S. Abbas and A. Abbas have recently expressed some criticisms (astro-ph/9612003) to my Letter "Biological Effects of Stellar Collapse Neutrinos" (J.I. Collar, Phys. Rev. Lett. **76** (1996) 999, astro-ph/9505028).

• The evident intent of that Letter is to examine the possible *contribution* of neutrinos from stellar collapse to the paleontological record of mass extinctions (verbatim, from the abstract). In view of the differences in intensity, geological signatures, range of species affected and variety of candidate explanations for each of the numerous extinction episodes, it would be simplistic to assign one single cause to all of them. This is clearly not my intention nor, in my opinion, that of the authors of refs. [1,2], to which Abbas & Abbas extend their criticism. However, Abbas & Abbas seemingly believe it is, and base all their arguments on this. To find, as in my Letter, that the frequency of a process is compatible with the apparent periodicity of major extinctions *is not* to claim that this process is therefore the origin of all of them, but that it takes place frequently enough to be a causing *candidate* in a possibly large fraction of them. The first conclusion is the self-serving point of departure for Abbas & Abbas, but is *nowhere* worded or implied in my Letter.

• Abbas & Abbas concentrate on the geological features of the K-T extinction in their attempt to refute my Letter. It is not without intention that this particular episode is *nowhere* mentioned in it, for I also happen to believe that there are compelling geological observations leading to ascribe the origin, in this case, to other causes. *Abbas & Abbas seem to forget that only in the last 250 Myr there has been eight extinction episodes well above background*. In this respect, their reaction is not different from that of some journalists who rushed to link my Letter to the demise of dinosaurs.

• Abbas & Abbas list a number of "fundamental and empirically established aspects" of extinctions, in their opinion at variance with the contents of my Letter and of refs. [1,2]. Yet these aspects are indeed far from being established:

- The question of periodicity, which they seem to take for granted, is not at all commonly accepted [3] and the limitations of this hypothesis are acknowledged by its authors [4] (for instance, this possible periodicity cannot be traced back into the Paleozoic).

 -Abbas & Abbas suggest (without providing any references) that an impact- or volcanism-induced iridium anomaly is a commonplace feature of significant extinctions. This is simply not so. It is well documented for a limited number of geographical sites in the case of the K-T episode, but, for instance, has not been observed elsewhere in the interval between the terminal Cretaceous to about 33 Myr ago [5,6], which includes the heavy extinction of the Eocene. While some much smaller excesses in the platinum-group elements have been reported coincidentally with extinction periods, explanations unrelated to accretionary episodes or volcanism have been favoured: redistribution of these elements by changes in sedimentary redox conditions [7], geochemical changes brought by regression of the sea or algae [8], changes from aerobic to anaerobic conditions in sediments or enhanced mid-ocean activity [9] and biological mechanisms [10]. Our present understanding of the low-temperature iridium geochemistry is too incomplete for an unambiguous interpretation to exist [7]. Finally, an iridium anomaly as large as the K-T boundary's is observed in the late Pliocene, a period during which no massive extinctions took place [11]. Their affirmation is, to say the least, misleading.

 - Abbas & Abbas claim that extinctions are believed to have occurred over extended periods of time. This is not true: the community is divided among those who believe in this, those who believe in sudden extinctions and a third group that tends to favour a mixed scenario [12]. A recent example is the work of Marshall and Ward [13,14], in which a new statistical technique points at the actual time of a specie's extinction using its last known fossil; this technique suggests abrupt extinctions.

 Not only the authors arrive to their own conclusions, extrapolating my paper to their convenience, but the arguments they use to refute them are tendentious, including some across-the-board affirmations about extinctions that might lead one to believe that everything is said and done in this field. Nothing is farther from reality.

In ref. [15] ("Clumpy Cold Dark Matter and Biological Extinctions", astro-ph/9512054) I suggested another possible extinction scenario: clumps of Weakly Interacting Massive Particles (WIMPs) are a natural prediction of certain cosmologies [16,17]; the passing of these clumps through the solar system would induce an unprecedented dose of high-LET radiation to organisms, in some cases comparable to a nearby nuclear detonation, yet affecting the whole biosphere. The duration of the crossings (~ few months) and their frequency are a function of WIMP mass. This frequency is again compatible in some interesting cases with the observed record of extinctions. The estimated high-LET radiation dose induced by this process can be much larger than in the stellar collapse scenario (provided that one is willing to accept the existence of such clumps!). It is hard to speculate about the outcome of a *global* exposure to Hiroshima neutrons at 1.5 km ground range, which is what I provide as a rough comparison: it seems plausible to believe that a large fraction of species would disappear due to radiation sickness while radiation-resistant organisms might suffer the effects of inheritable mutations.

In another recent pre-print, astro-ph/9612214 ("Volcanogenic Dark Matter and Mass Extinctions"), Abbas & Abbas depart from my paper and calculate the amount of heat produced by trapping and annihilation of these WIMPs in the Earth's core during the crossing of the clump. They find that this is the equivalent of up to ~10,000 times the annual heat flux in the Earth. Abbas & Abbas consider that this heat will be most likely dissipated in the form of slowly ascending plumes that ultimately result in explosive silicic volcanism followed by basalt volcanism. *However, in their own estimate, there is a delay of ~ 5 million years between the clump crossing and the arrival of these plumes to the Earth's surface*. Abbas & Abbas propose that this is a possible explanation for the large-scale volcanism that may have accompanied several extinction episodes.

At this point Abbas & Abbas stop to mention that the hypothesis put forward in my paper (extinctions due to high-LET radiation during the crossing) fails to explain the geological signatures characteristic of large-scale volcanism, in what appears to be an attempt to disqualify the crossing itself as the origin of extinctions (?). Yet, they do not try to refute the estimated severity of the WIMP-induced radiation dose. In view of the ~5 million year delay between radiation and volcanism, this is a surprising breach in logic. While their pre-print is not devoid of merit, in having found a possible geological *signature* for clump crossing, it is logical to expect that the belated volcanic fireworks might find few witnesses or that, at the most, they would be a *coup de grâce* to the surviving species.

J. I. Collar
PPE Division
CERN
Geneve 23, CH-1211